\documentclass{blois}

\usepackage[superscript]{cite}
\usepackage{lineno}
\usepackage{comment}

\def\be{\begin{equation}}
\def\ee{\end{equation}}
\def\bea{\begin{eqnarray}}
\def\eea{\end{eqnarray}}

\newcommand{\Photo}{}

\begin{document}
\vspace*{4cm}
\title{HIGHLIGHTS ON SUPERSYMMETRY AND EXOTIC SEARCHES \\ AT THE LHC~\footnote{Copyright 2022 CERN for the benefit of the ATLAS, CMS and LHCb Collaboration. CC-BY-4.0 license}}

\author{ S. SEKMEN \\ on behalf of the ATLAS, CMS and LHCb Collaborations }

\address{Center for High Energy Physics, Kyungpook National University \\
Daegu, South Korea}

\maketitle\abstracts{
The Run 2 data taking period of the Large Hadron Collider (LHC) at CERN in years 2015-2018 has presented a great opportunity to search for physics beyond the standard model (BSM).  It will be followed by the Run 3 period starting in 2022, and by the High-Luminosity LHC (HL-LHC) era starting in late 2020s, where the latter promises an unprecedented wealth of physics prospects due to very high expected integrated luminosity and improved detector features.  The ATLAS, CMS and LHCb experiments pursued a rich physics program in Run 2, and are already assessing the physics expectations at the HL-LHC era.  This report presents highlights from recent search results and HL-LHC studies on BSM physics by the ATLAS, CMS and LHCb experiments.  Examples will be shown from model-independent generic searches, searches for supersymmetry, extended Higgs sectors, and new exotic fermions and bosons.}

\section{Introduction}

The Large Hadron Collider (LHC) at CERN successfully completed its Run 2 operations between 2015-2018, during which  ATLAS~\cite{ATLAS:2008xda}, CMS~\cite{CMS:2008xjf} and LHCb~\cite{LHCb:2008vvz} experiments collected up to $\sim$140~fb$^{-1}$ of proton-proton collision data at 13 TeV center of mass energy. These data were explored in a rich physics program, resulting in hundreds of searches for hints of beyond the standard model (BSM) physics.  Run 2 searches for new particles and new interactions targeted the most challenging and interesting signatures. They employed highly innovative analysis methods, where machine learning played an increasingly dominant role.  The searches also made more refined use of detector capabilities in physics object reconstruction and identification. For example, objects with high Lorentz boost were identified with improved jet substructure techniques and were used more extensively.  Moreover, a diverse set of studies were designed featuring final states with long-lived particles.

LHC is scheduled to start Run 3 operations in 2022.  This will be followed by an extensive period of so-called Phase-2 upgrades for the accelerator and the detectors in preparation for the High-Luminosity LHC (HL-LHC) era scheduled to start in late 2020s.  HL-LHC will operate at 14 TeV energy and collect 3 ab$^{-1}$ of data.  It will offer extended physics opportunities resulting from more data, increased detector coverages and new detector features.  Many studies are already performed to explore the full potential of the HL-LHC and upgraded detectors related to BSM physics, including BSM extensions of the Higgs sector~\cite{CidVidal:2018eel, Cepeda:2019klc}. 

BSM searches at the LHC share several characteristics.  Searches are usually signature-based, meaning that they are designed to explore generic final states (e.g. dileptons, multitops, ...), rather than targeting specific BSM models.  A search in a certain final state often probes multiple BSM scenarios.  In some cases, model-independent results are presented for reinterpretation purposes.  Run 2 BSM searches focus on cases with increasingly smaller signals and larger backgrounds.  The analysis methodology typically involves applying an event selection to enhance signal and eliminate backgrounds (which involves increased use of machine learning discriminants in Run 2) and estimating the backgrounds via data control regions or Monte Carlo simulations. A blind analysis strategy is adopted, where the analysis methodology is thoroughly validated before comparing data with estimated background in search regions.  Finally, sensitivity to predicted theoretical quantities (such as cross sections, masses) is quoted, taking into account the effects of uncertainties from theoretical and experimental sources.

This report presents highlights from recent BSM search results from Run 2 (along with a few relevant results from the earlier Run 1) reported by the ATLAS, CMS and LHCb experiments.  Projections for BSM physics at the HL-LHC are also summarized.  Examples are presented from model-independent generic searches, searches for supersymmetry, extended Higgs sectors, new exotic fermions and new exotic bosons.

\section{Model-independent generic searches}

\begin{figure}[htp!]
\centering
$\vcenter{\hbox{\includegraphics[width=0.45\textwidth]{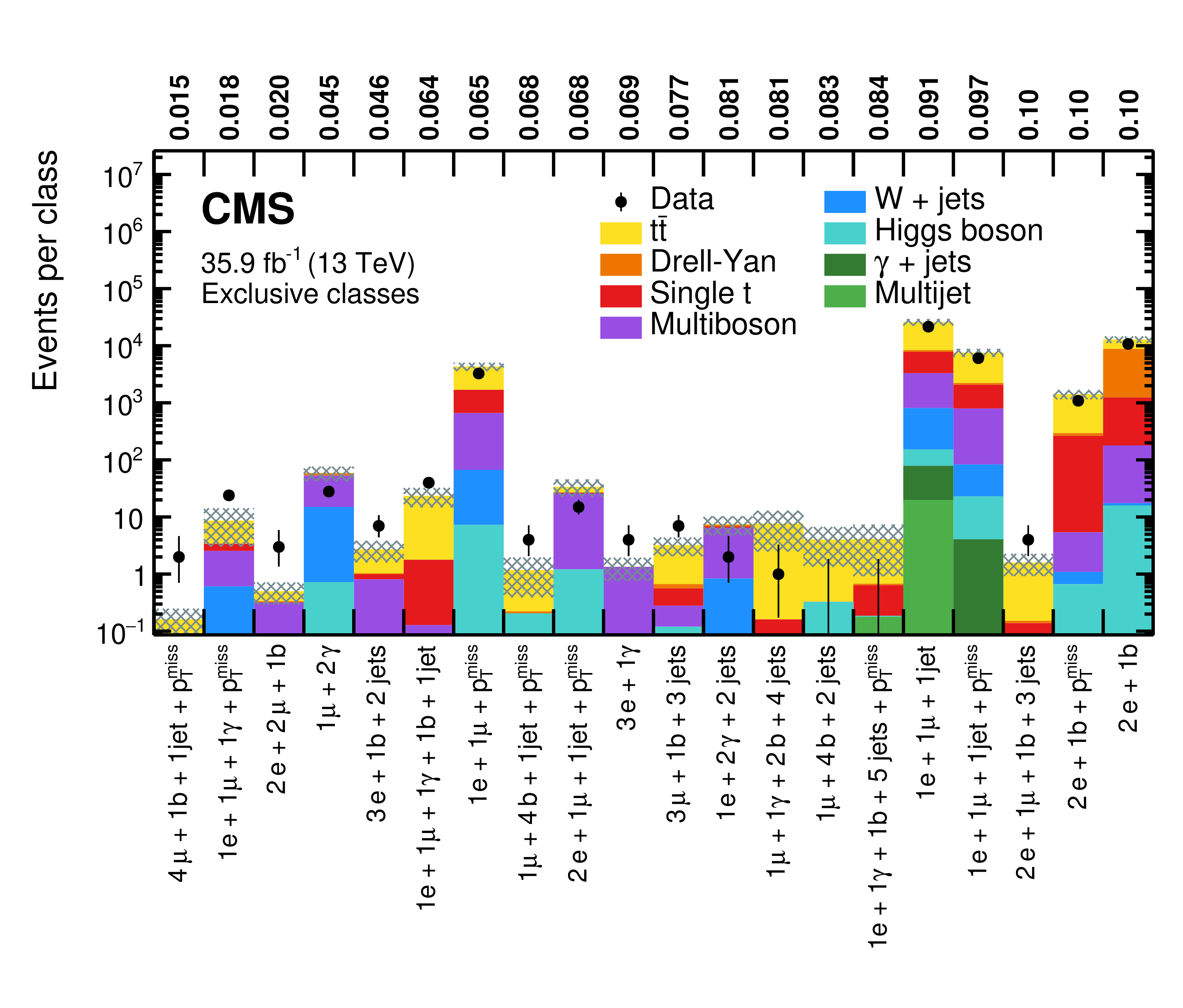}}}$
\hspace{0.8cm}
$\vcenter{\hbox{\includegraphics[width=0.40\textwidth]{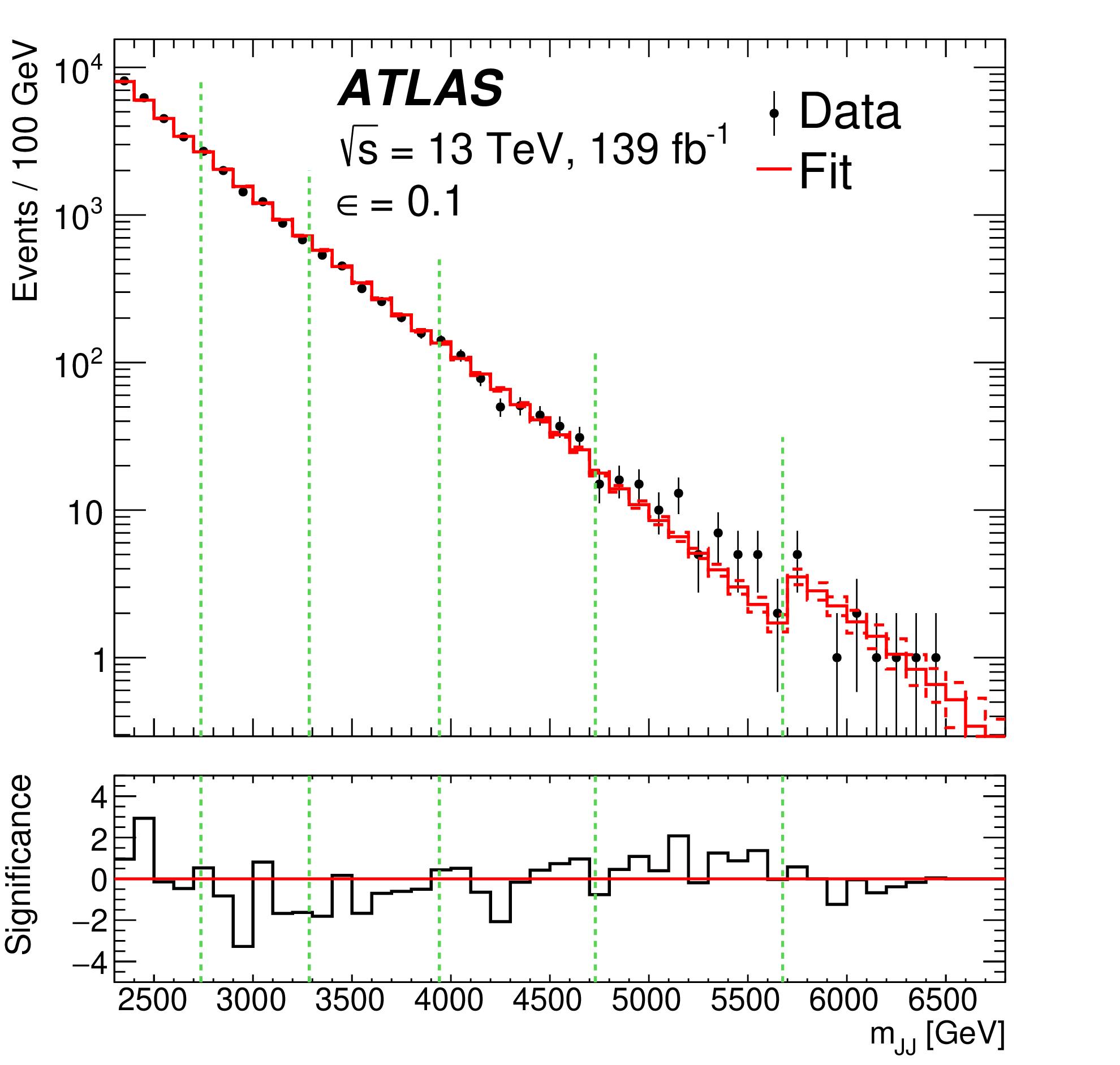}}}$
\caption{Left: Data and SM predictions for the most significant exclusive event classes, where the significance of an event class is calculated in a single aggregated bin from the CMS Run 2 MUSIC search~\cite{CMS:2020zjg}.  Right: A comparison of the fitted background and the data in all six considered signal regions from the ATLAS Run 2 CwoLA search~\cite{ATLAS:2020iwa}.
\label{fig:genericsearches}}
\end{figure}

Both ATLAS and CMS perform generic and inclusive searches looking for excesses over the standard model (SM) expectations simultaneously in multiple final states or extended regions of kinematic phase space.  One example is the Model Unspecific Search in CMS (MUSIC) which looks for discrepancies between observed data and SM expectations in about 1000 different final states defined by different physics object multiplicity content~\cite{CMS:2020zjg}.  The search scanned for deviations in scalar sum of object transverse momenta, invariant masses, transverse masses and missing transverse energy ($E_T^{miss}$), and found no significant deviations from the SM.  Figure~\ref{fig:genericsearches} (left)  shows event classes observed with highest significance in aggregated search bins. 
Another study titled Classification without Labels (CwoLA) is a search for resonant new physics using a machine learning-based anomaly detection procedure that does not rely on a signal model hypothesis~\cite{ATLAS:2020iwa}. Weakly supervised learning was used to train classifiers directly on data to enhance potential signals. Analysis targeted dijet event topologies and  the masses of the two jets were used as input features for the machine learning algorithm. The analysis is a three-dimensional search $A\rightarrow BC$, for $m_A \sim $O(TeV), $m_B,m_C \sim$ O(100 GeV), where $B$, $C$ were reconstructed as large-radius jets.    Figure~\ref{fig:genericsearches} (right) shows a comparison of the fitted background and the data in all six considered signal regions. The analysis observed no localized excess in dijet invariant mass from 1.8 to 8.2 TeV.

\section{Supersymmetry searches}

Supersymmetry (SUSY) is a well-motivated theoretical framework offering solutions to deficiencies of the SM. It proposes that every SM particle has a heavier superpartner with different spin.  In its most generic representation, SUSY has more than $~\sim$100 free parameters, leading to varying values for sparticle masses, cross sections and branching ratios.  ATLAS and CMS perform many searches targeting different flavors of SUSY and sparticle mass spectra.  In the R-parity conserving SUSY models, sparticles are produced in pairs and the lightest SUSY particle (LSP), which is usually a neutralino and a dark matter (DM) candidate, is stable.  These scenarios lead to final states with high $E_T^{miss}$, high object multiplicities and objects high visible transverse momentum ($p_T$).  In the R-parity violating SUSY models, LSP is unstable and decays to SM particles, resulting in the absence of genuine $E_T^{miss}$.  For compressed SUSY scenarios with small mass differences between sparticles, final states have both low $E_T^{miss}$ and objects with low $p_T$.  SUSY searches are typically interpreted using simplified models (SMS), which are effective Lagrangian descriptions defined by sparticle masses, along with production and decay processes.  A single search is usually interpreted with multiple SMSs.

\begin{figure}[htp!]
\centering
$\vcenter{\hbox{\includegraphics[width=0.46\textwidth]{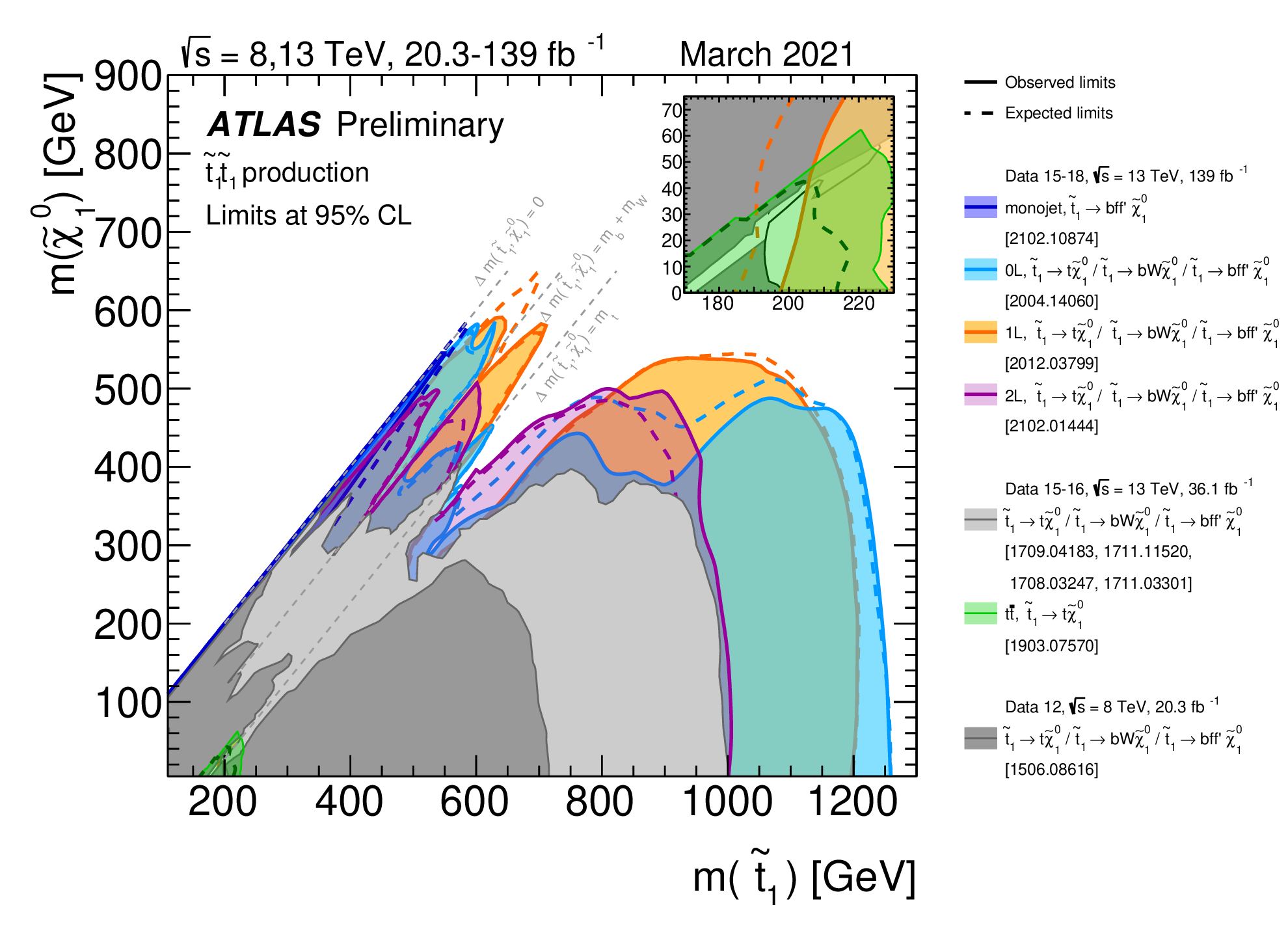}}}$
\hspace{0.5cm}
$\vcenter{\hbox{\includegraphics[width=0.35\textwidth]{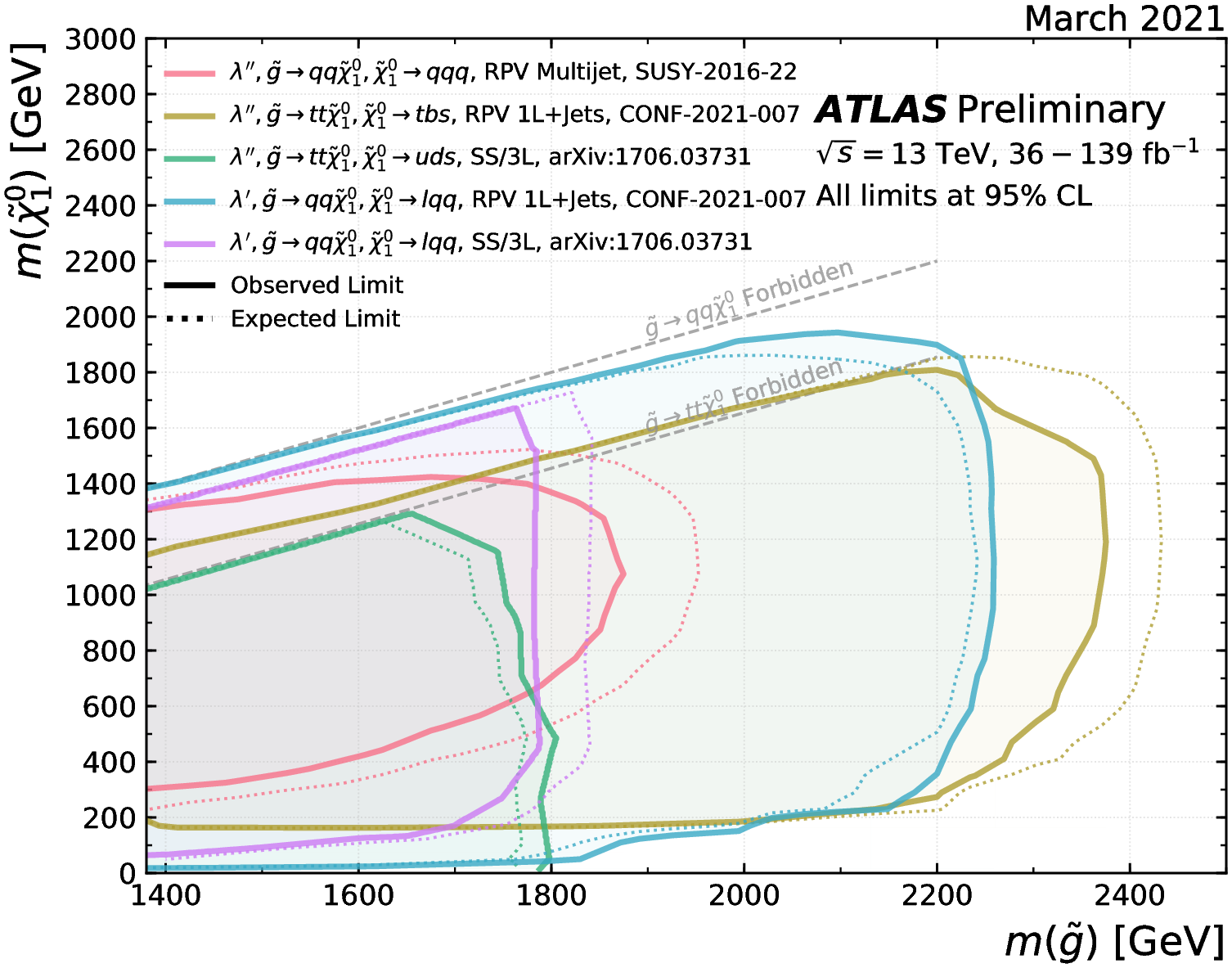}}}$
\caption{Left: Summary of the dedicated ATLAS searches for top squark pair production based on Run 2 data.  Exclusion limits at 95\% CL are shown in the top squark-neutralino mass plane.  Different decay channels have been considered~\cite{ATL-PHYS-PUB-2021-019}.  Right: ATLAS exclusion limits for RPV light squarks on the gluino-LSP mass plane from different searches~\cite{ATL-PHYS-PUB-2021-019}.}
\label{fig:susy1}
\end{figure}

Gluinos and squarks have high production cross sections and decay through multiple channels. They are explored mainly via inclusive searches in final states with multiple jets, 1,2 leptons, 1,2 photons, etc, that are sensitive to different decay channels. Multiple disjoint search regions defined by object multiplicities and kinematic variables enhance sensitivity for different mass configurations.  Different searches (e.g.~\cite{ATLAS:2020syg, ATLAS-CONF-2018-041, ATLAS:2021twp, CMS:2019zmd, CMS:2019ybf, CMS:2020cur}) have probed gluino masses up to 2.4 TeV, light squark masses up to 1.8 TeV and bottom squark masses up to 1.3 TeV.  Searches for top squarks are particularly important since the Naturalness hypothesis implies their existence at masses around 1-2 TeV.  Dedicated searches are designed to probe different regions of the top squark-neutralino mass plane. ATLAS exclusion limits from recent searches (e.g.~\cite{ATLAS:2021kxv, ATLAS:2020dsf, ATLAS:2020xzu, ATLAS:2021hza}) are shown in Figure~\ref{fig:susy1} (left), where the highest exclusion is achieved by the zero lepton searches, and reaches up to top squark masses of 1.25 TeV.  A combination of CMS searches in 0, 1 and 2 lepton channels excludes top squarks up to 1.35 TeV~\cite{CMS:2021eha}.  A simple ATLAS study projected the top squark exclusion at HL-LHC to extend to 1.7 TeV~\cite{CidVidal:2018eel}.  Searches for RPV gluinos and squarks have also been performed in final states with high object multiplicity and no $E_T^{miss}$.  These searches look for resonances, which would appear as bumps in the invariant mass distributions.  Figure~\ref{fig:susy1} (right) shows the ATLAS exclusion limits for RPV light squarks on the gluino-LSP mass plane from different searches, e.g.~\cite{ATLAS:2021fbt}.

\begin{figure}[htp!]
\centering
$\vcenter{\hbox{\includegraphics[width=0.35\textwidth]{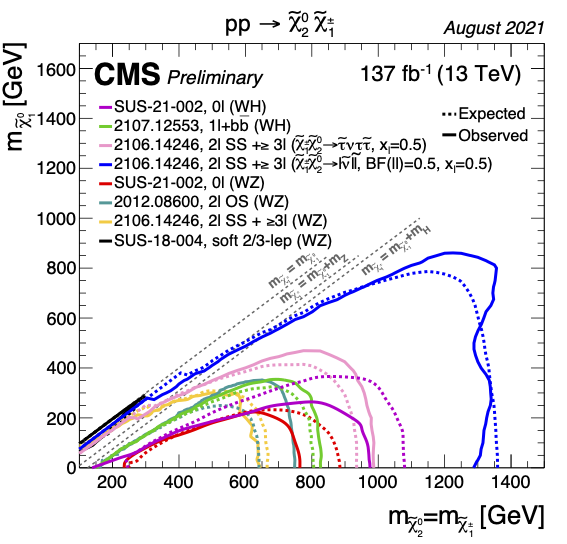}}}$
\hspace{0.5cm}
$\vcenter{\hbox{\includegraphics[width=0.46\textwidth]{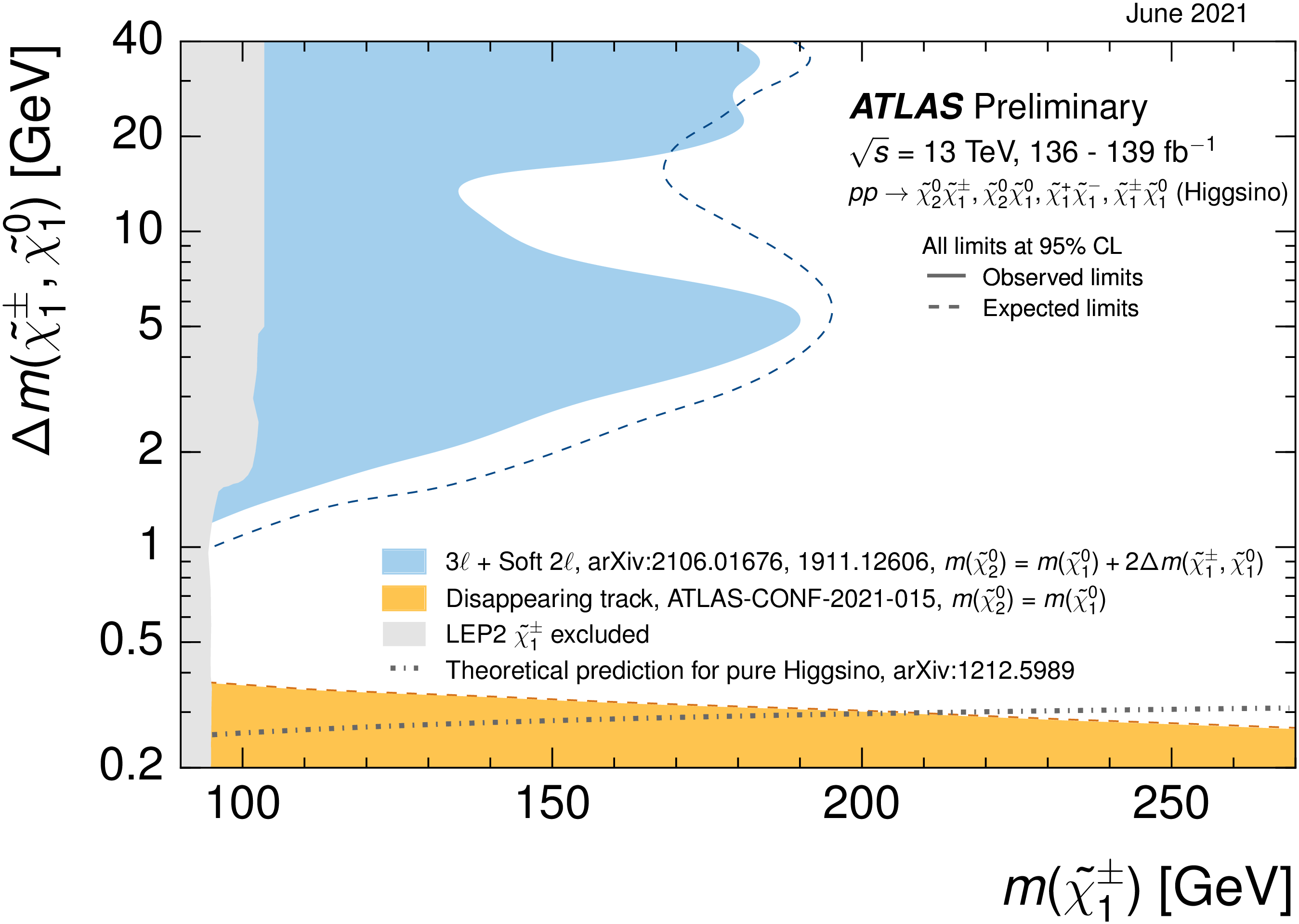}}}$
\caption{Left: Summary of the dedicated CMS searches for $\tilde{\chi}^0_2 \tilde{\chi}^\pm_1$ production based on Run 2 data.  Exclusion limits at 95\% CL are shown  in the $\tilde{\chi}^\pm_1/\tilde{\chi}^0_2$-$\tilde{\chi}^0_1$ mass plane.  Different decay channels have been considered.  
Right: Exclusion limits at 95 CL for higgsino pair production $\tilde{\chi}_1^+\tilde{\chi}_1^-$, $\tilde{\chi}_1^\pm\tilde{\chi}_1^0$, $\tilde{\chi}_1^\pm\tilde{\chi}_2^0$ and $\tilde{\chi}_2^0\tilde{\chi}_1^0$ with off-shell SM-boson-mediated decays to $\tilde{\chi}^0_1$, as a function of the $\tilde{\chi}^\pm_1$ and $\tilde{\chi}^0_1$ masses from ATLAS Run 2 soft dilepton and disappearing track searches. The production cross-section is for pure higgsinos~\cite{ATL-PHYS-PUB-2021-019}. }
\label{fig:susy2}
\end{figure}

Direct production of neutralinos and charginos ($\tilde{\chi}^0$ and $\tilde{\chi}^\pm$) have lower cross sections, where the cross section depends on electroweak state composition of the mass eigenstates.  They can decay in a variety of ways, either directly via leptons, or indirectly via sleptons, sneutrinos or W/Z/h bosons.  Searches for direct production of charginos and neutralinos exploit final states with multileptons, and more recently with Lorentz-boosted hadronic W/Z/h bosons.  Figure~\ref{fig:susy2} (left) shows exclusion limits for $\tilde{\chi}^0_2 \tilde{\chi}^\pm_1$ production in the $\tilde{\chi}^\pm_1/\tilde{\chi}^0_2$-$\tilde{\chi}^0_1$ mass plane from several recent CMS analyses~\cite{CMS-PAS-SUS-21-002, CMS:2021few, CMS:2021cox}.  The searches were able to probe $\tilde{\chi}^\pm_1/\tilde{\chi}^0_2$ masses up to 1.35 TeV. 

Charginos and neutralino mass eigenstates are composed of electroweak eigenstates of binos, winos and higgsinos, i.e. the superpartners of the weak hypercharge gauge boson, of the weak SU(2) gauge bosons, and of the Higgs bosons,  respectively.  Naturalness argument predicts the existence of higgsinos (higgsino dominated $\tilde{\chi}^\pm$, $\tilde{\chi}^0$) below 1 TeV.  Higgsinos have smaller direct production cross sections with respect to the binos or winos, and they lead to compressed final states with small mass differences $\Delta m(\tilde{\chi}^\pm_1, \tilde{\chi}^0_1)$.  
For larger $\Delta m$, $\tilde{\chi}^\pm_1$ decays promptly, leading to final states with low-$p_T$ (soft) leptons and $E_T^{miss}$. Both ATLAS and CMS studied these final states in~\cite{ATLAS:2021moa, CMS:2021edw}.  For smaller $\Delta m$, $\tilde{\chi}^\pm_1$ becomes long-lived and usually decays within the tracker volume to very soft, undetectable pions and a $\tilde{\chi}^0_1$, leaving a short track with missing outer hits.  This leads to a so-called ``disappearing track" $+ E_T^{miss}$ signature, investigated by both ATLAS and CMS~\cite{ATLAS:2022rme, CMS:2020atg}.  Figure~\ref{fig:susy2} (right) shows the regions excluded by the ATLAS searches in $\Delta m(\tilde{\chi}^\pm_1, \tilde{\chi}^0_1)$ versus $m(\tilde{\chi}^\pm_1)$, displaying the complementarity of the two analyses.  At the HL-LHC, sensitivity will improve due to new tracking detectors.  

Besides those explained above, an increasing multitude of dedicated searches are being designed to explore more distant and unique corners of the SUSY parameter space. These include searches with Higgs boson in cascade decays, searches for vector-boson fusion production, searches with photons, with long-lived particles and searches for extended SUSY models with extra particles.  One example is a CMS search for stealth SUSY, which is a scenario extending the minimal supersymmetric standard model (MSSM) with a hidden sector.  The CMS search looked for two stealth top squarks in a final state of 1 lepton, $\ge 7$ jets and no $E_T^{miss}$, using neural networks, and excluded stealth top squarks up to mass 870 GeV~\cite{CMS:2021knz}.

\section{Searches for extended Higgs sectors}

\begin{figure}[htp!]
\centering
$\vcenter{\hbox{\includegraphics[width=0.48\textwidth]{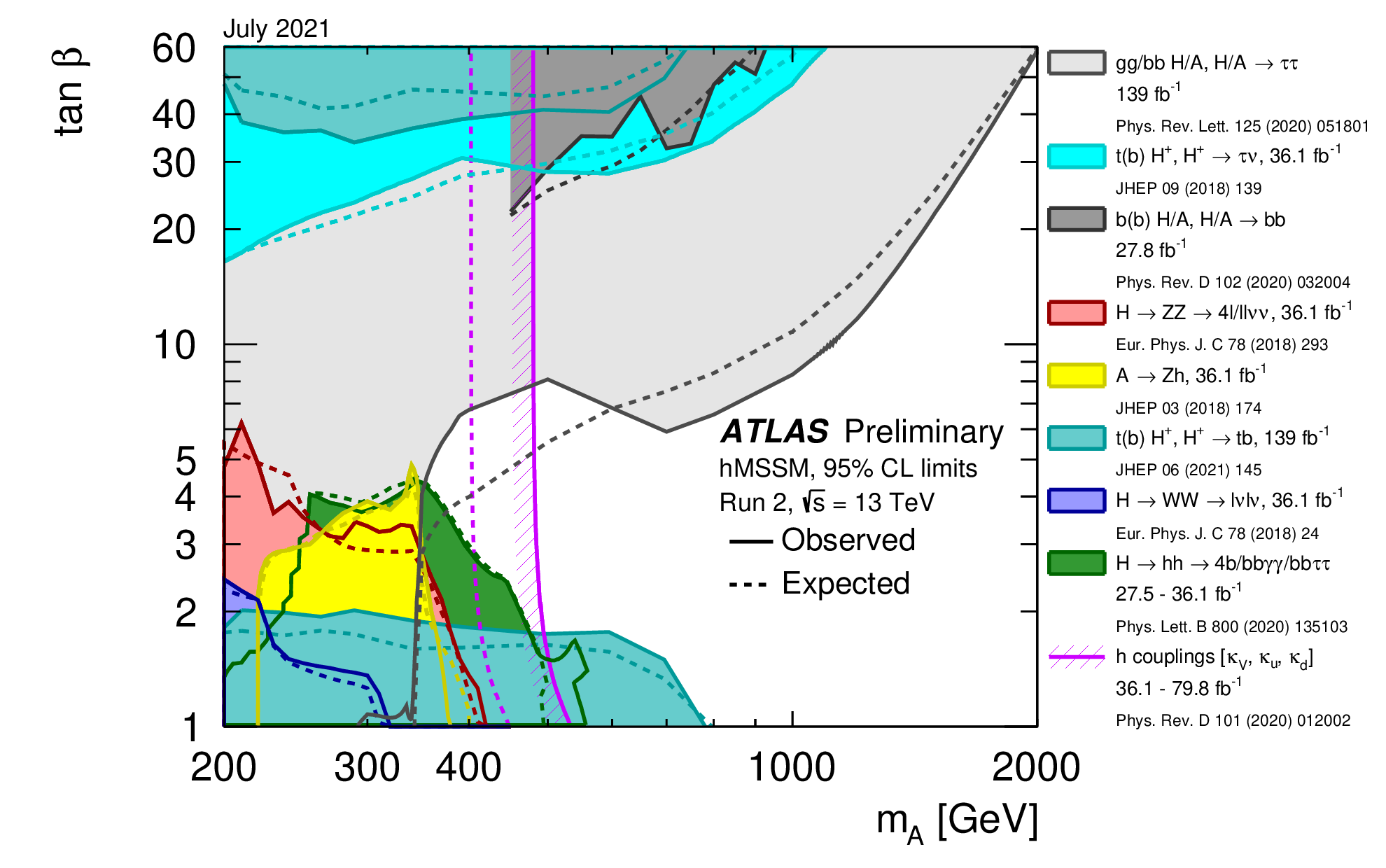}}}$
\hspace{0.5cm}
$\vcenter{\hbox{\includegraphics[width=0.35\textwidth]{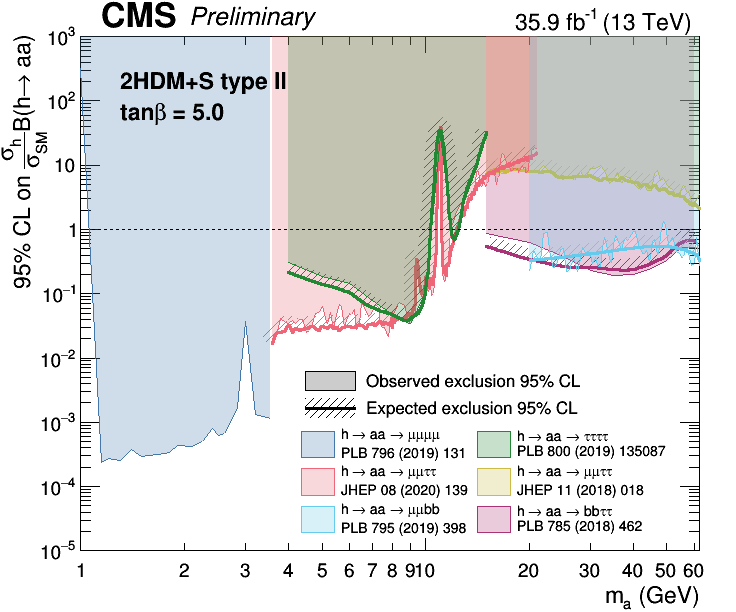}}}$
\caption{Left: 95\% CL exclusion limits from several ATLAS Run 2 searches in the [$m_A, \tan\beta]$ plane for hMSSM via direct searches for heavy Higgs bosons and fits to the measured rates of observed Higgs boson production and decays~\cite{ATL-PHYS-PUB-2021-030}.  Right: 95\% CL exclusion limits on $\frac{\sigma(h)}{\sigma_{SM}} \times BR(h \rightarrow aa)$ in the 2HDM+S Type-II model with $\tan \beta = 5$ for exotic h decay searches performed by CMS in Run 2. }
\label{fig:2HDM}
\end{figure}

Some BSM theories suggest extending the Higgs sector with extra fields and Higgs particles.  These extended Higgs sectors are explored by an extensive set of searches by ATLAS and CMS.  One case is the Two Higgs Doublet Model (2HDM) where SM is extended by two Higgs doublets, which can couple to SM fermions in several different ways. The 2HDM Higgs sector consists of 5 physical Higgs bosons: two CP-even neutral Higgs bosons $h$ (which is equivalent to the 125 GeV SM-like boson observed at the LHC) and the heavier $H$; two charged Higgs bosons $H^\pm$; and one CP-odd neutral Higgs boson $A$.  The MSSM is a Type-II 2HDM, where one doublet couples to up-type fermions while the other couples to down-type fermions.  Here, the Higgs sector is determined at tree level by two parameters, namely the mass of the pseudoscalar $A$ ($m_A$) and the ratio of the vacuum expectation values of the two doublets ($\tan\beta$).  Figure~\ref{fig:2HDM} (left) shows the regions of the [$m_A$ , $\tan\beta$] plane excluded in the so-called hMSSM submodel through several direct ATLAS searches for heavy Higgs bosons~\cite{ATL-PHYS-PUB-2021-030}.   Searches for the extra Higgs particles typically explore Higgs decays to SM particles.  For decays to visible final states, searches look for excesses in invariant mass distributions, while for decays to visible particles and neutrinos, searches look for excesses in transverse mass distributions.  HL-LHC projections for $H/A\rightarrow \tau\tau$ were also performed and can be found in~\cite{Cepeda:2019klc}.

Two Higgs doublet models can be further extended by the addition of a singlet field.  This case is realized in the well-motivated Next-to-Minimal Supersymmetric Standard Model (NMSSM) scenarios.  The NMSSM Higgs sector consists of 7 physical Higgs bosons, namely the CP-even $h_1 \equiv h$ (SM-like), $h_2$ and $H_3$; the CP-odd $a_1 \equiv a$ and $A_2$; and the charged $H^\pm$.  Decays of $h_1 \rightarrow aa$ are possible. For low $m_a$, the pseudoscalar $a$ would have a high Lorentz boost and lead to merged decay products.  Searches are designed to investigate $h \rightarrow aa$ decays, varying $m_a$ up to $m_{h}/2$, where for $aa \rightarrow \mathrm{visible}$ decays, they look for an excess in the 4 object invariant mass and for $aa \rightarrow \mathrm{visible + neutrinos}$ decays, they look for an excess in a transverse mass computed from the visible and invisible decay products.  Figure~\ref{fig:2HDM} (right)  shows the exclusion limits on $\frac{\sigma(h)}{\sigma_{SM}} \times BR(h \rightarrow aa)$ in the 2HDM+S Type-II model with $\tan \beta = 5$ for exotic h decay searches performed by CMS in Run 2 (most recent being~\cite{CMS:2020ffa}).  CMS also performed an HL-LHC search for $aa \rightarrow bb\tau\tau$ and $aa \rightarrow \mu\mu\tau\tau$ in final states with hadronic and leptonic $\tau$ decays~\cite{Cepeda:2019klc}. The analysis is expected to be sensitive to effective signal strength $\frac{\sigma(h)}{\sigma(h_{SM})}\times BR(h \rightarrow aa \rightarrow b\bar{b}\tau\tau/ \mu\mu\tau\tau)$ values down to 1.  

\begin{figure}[htp!]
\centering
$\vcenter{\hbox{\includegraphics[width=0.36\textwidth]{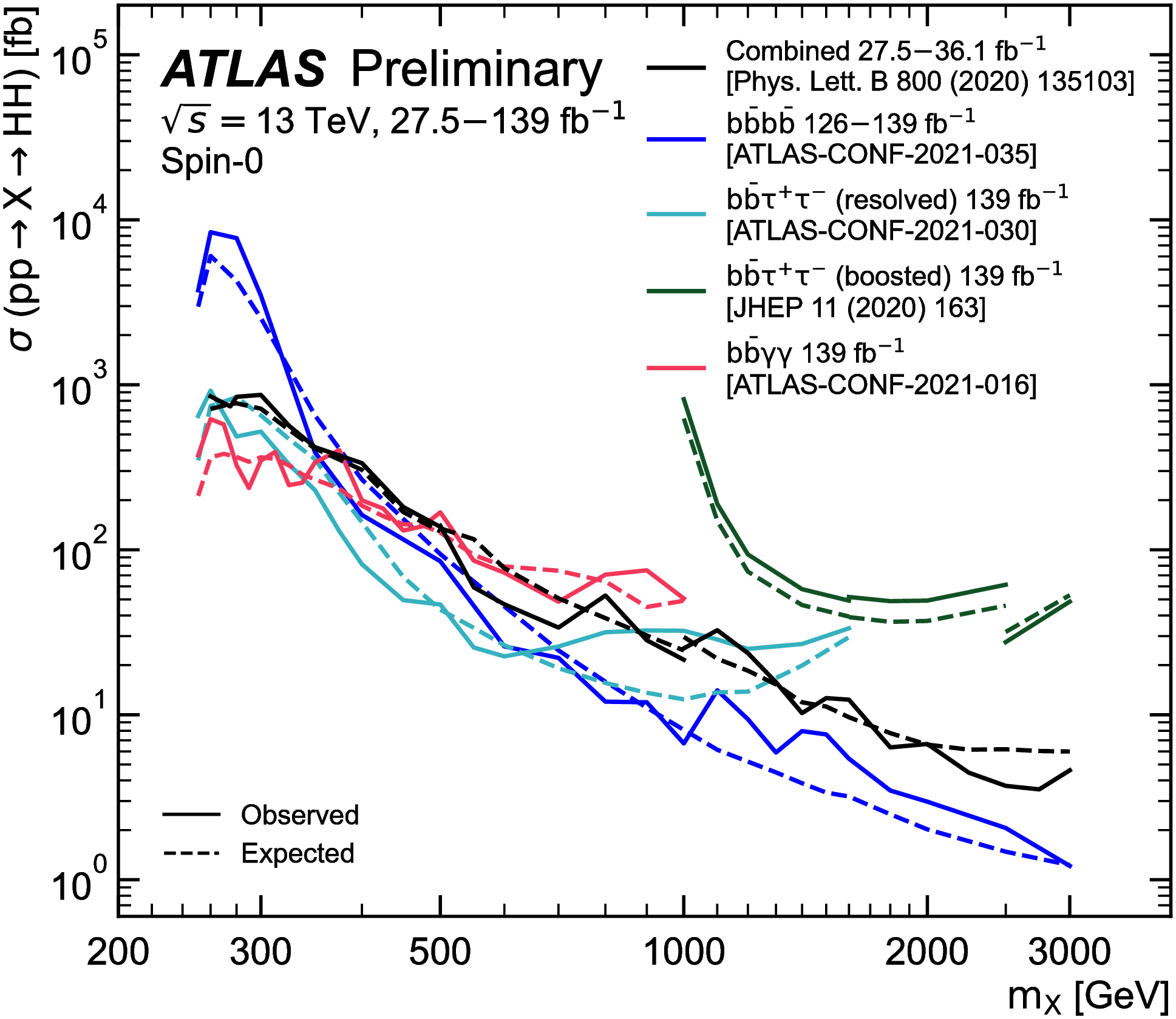}}}$
\hspace{1cm}
$\vcenter{\hbox{\includegraphics[width=0.36\textwidth]{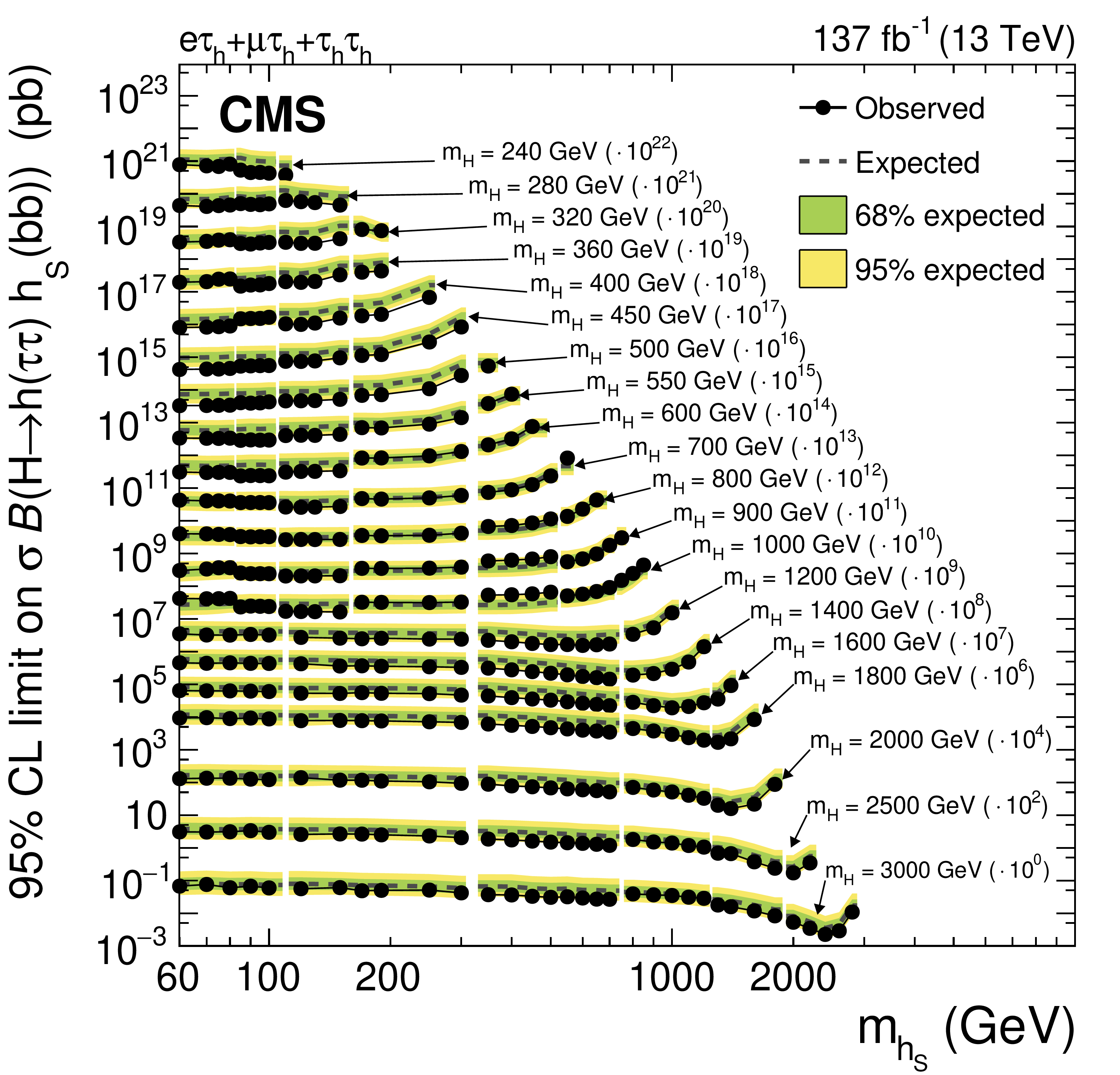}}}$
\caption{Left: Upper limits at 95\% CL from ATLAS searches on the resonant HH production cross-section as a function of the mass for a narrow-width scalar resonance~\cite{ATL-PHYS-PUB-2021-031}. Right: Expected and observed 95\% CL upper limits on $\sigma \times BR(H \rightarrow h(\tau\tau) h_s(b\bar{b}))$ for different values of $m_H$ and $m_{h_S}$~\cite{CMS:2021yci}.}
\label{fig:HH}
\end{figure}

A further probe of Higgs -- BSM interplay is the resonant di-Higgs production ($X \rightarrow hh$), where examples of the resonance could be a heavy BSM particle such as a generic spin-0 boson or a Kaluza-Klein graviton.  Searches are performed in both gluon-gluon fusion (ggF) and vector boson fusion (VBF) production channels, in various combinations of the Higgs boson decay products, such as $b\bar{b}b\bar{b}$, $b\bar{b}\tau^+\tau^-$ and $b\bar{b}\gamma\gamma$.  Figure~\ref{fig:HH} (left) shows 95\% CL upper limits from ATLAS searches on the resonant $hh$ production cross-section as a function of the mass for a narrow-width scalar resonance $X$ (including~\cite{ATLAS:2022hwc, ATLAS-CONF-2021-030, ATLAS:2021ifb}).  Another search performed by CMS looked for a heavy Higgs boson H decaying into the observed SM-like Higgs boson ($h_{125}$) and a new light Higgs boson in the 2HDM$+$singlet models ($h_S$), where the $h_{125}$ and $h_S$ are required to decay into a pair of tau leptons and a pair of b quarks, respectively~\cite{CMS:2021yci}.  No signal has been observed within mass ranges of 240-3000 GeV for $H$ and 60-2800 GeV for $h_S$.  Neural networks were used for signal classification.  Figure~\ref{fig:HH} (right) shows the model independent 95\% CL upper limits on the product of the production cross section and the branching ratios ($\mathrm{\sigma \times BR}$) of the signal process. These limits have been compared to maximally allowed $\mathrm{\sigma \times BR}$ of the signal processes predicted by the NMSSM. 

Dedicated searches are also performed to look for charged Higgs bosons. One recent example is a CMS search for the VBF production of mass-degenerate charged and doubly-charged Higgses $H^\pm$ and $H^{\pm\pm}$, decaying into vector bosons~\cite{CMS:2021wlt}. The search was performed in final states with 2 same charge leptons or 3 leptons plus 2 jets with large rapidity separation and large invariant mass.  A maximum likelihood fit was performed to the reconstructed dijet invariant mass ($m_{jj}$) and a transverse mass calculated for the di-vector boson system ($m_T^{VV}$) to extract signal, where no deviations from the SM are observed.  The results were interpreted in the Georgi-Machacek (GM) model where the observed 95\% CL limits exclude GM $s_H$ parameter values greater than 0.20 to 0.35 for the $H^{\pm}, H^{\pm\pm}$ mass range from 200 to 1500 GeV.  Additionally, ATLAS studied the pair production $H^{++}H^{--}$ and the associated production $H^{\pm\pm}H^{\mp}$ for $H^{\pm\pm} \rightarrow W^\pm W^\pm$ and $H^{\pm} \rightarrow W^\pm Z$ in final states with 2 same charge leptons, or 3 or 4 leptons with a variety of charge combinations, jets and $E_T^{miss}$~\cite{ATLAS:2021jol}.  No significant deviations from the SM predictions were observed. The results are interpreted in terms of the Type-II seesaw model that extends the SM scalar sector with a scalar triplet.  $H^{\pm\pm}$ bosons were excluded at 95\% CL up to 350 GeV and 230 GeV for the pair and associated production modes, respectively.

\section{Searches for new exotic fermions and bosons}

Many candidate models for BSM physics predict the existence of new fermionic or bosonic particles, and a multitude of searches are designed to look for these new particles in a variety of final states. The first example is the excited states of quarks and leptons predicted by the compositeness models, where leptons and quarks are composite objects made of more fundamental constituents. Compositeness models postulate new interactions that occur above a compositeness scale $\Lambda$.   CMS searched for excited b quarks, $b^*$, by looking for dijet resonances with at least one jet coming from a b quark~\cite{CMS-PAS-EXO-20-008}.  Energetic b quarks in jets were identified by a deep neural network.  The analysis looked for excess in dijet invariant mass over a fit to the SM background prediction.  No excess was observed and excited $b^*$ were excluded up to mass 4 TeV. 

\begin{figure}[htp!]
\centering
$\vcenter{\hbox{\includegraphics[width=0.52\textwidth]{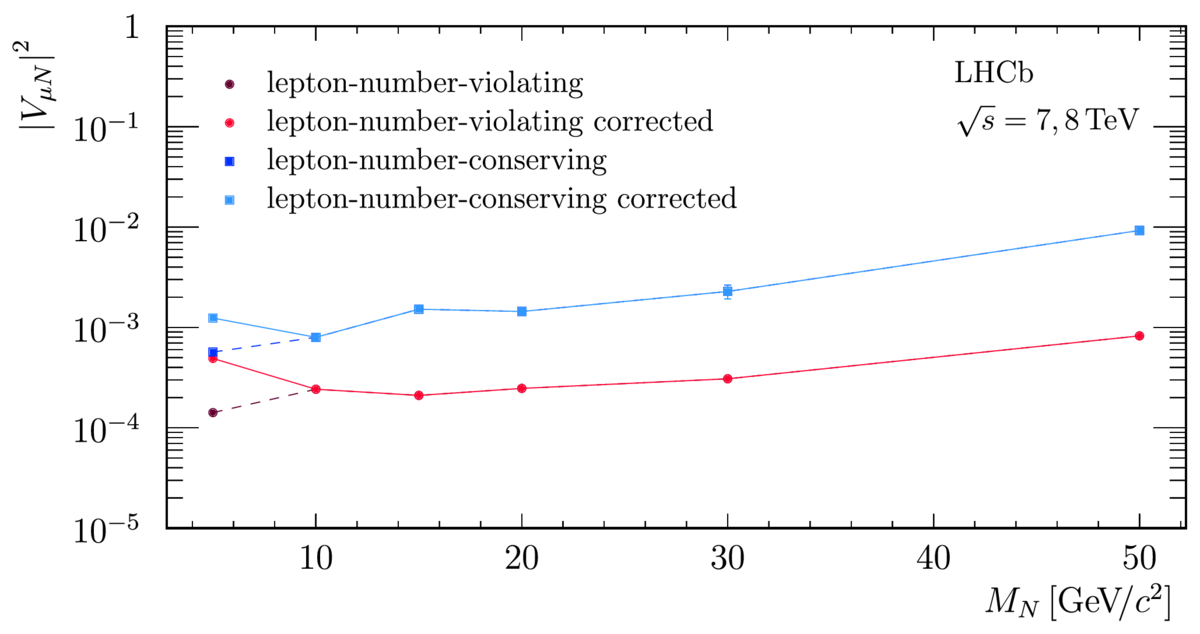}}}$
\hspace{0.5cm}
$\vcenter{\hbox{\includegraphics[width=0.37\textwidth]{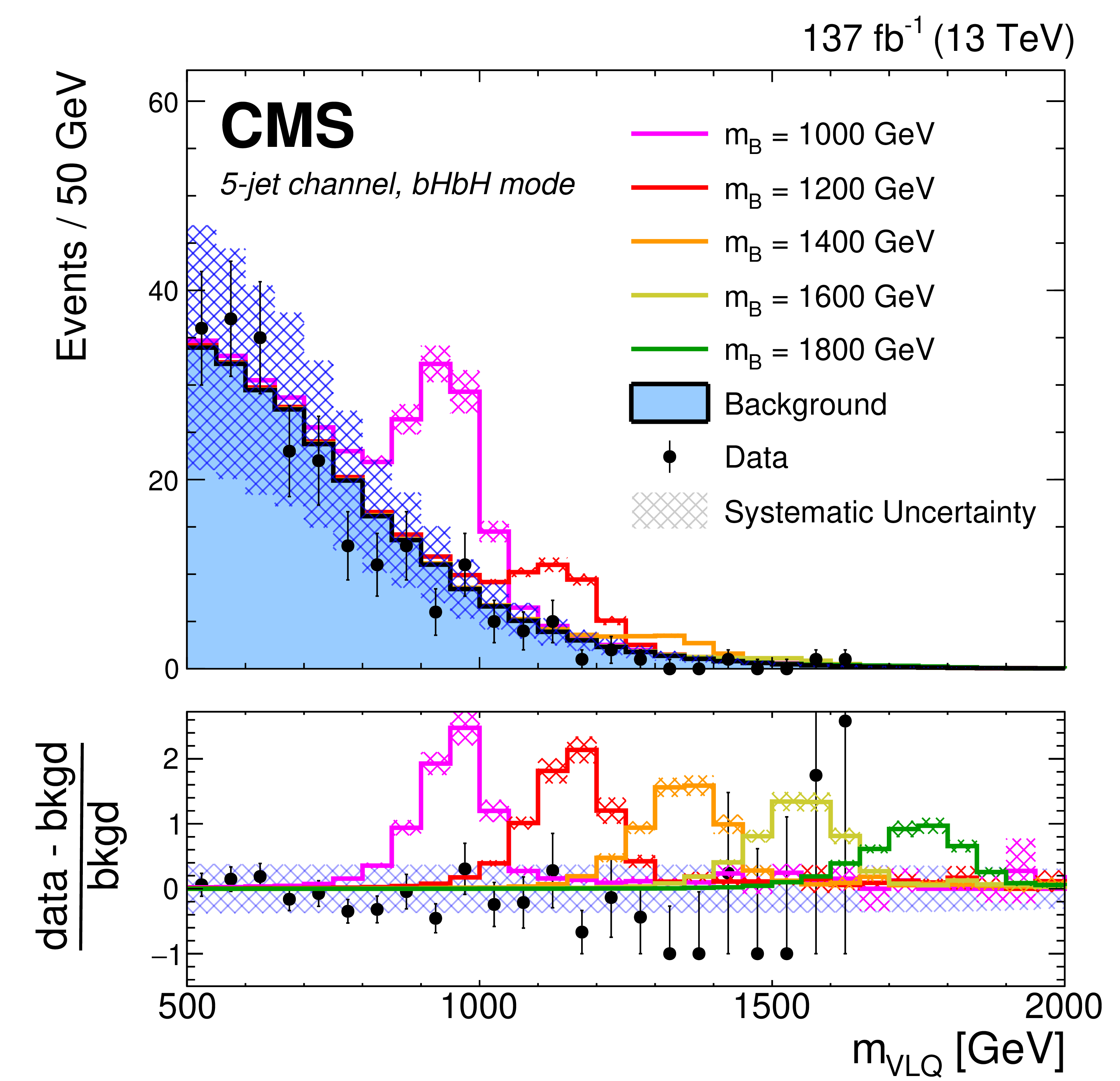}}}$
%$\vcenter{\hbox{\includegraphics[width=0.43\textwidth]{figures/LHCb_LFVh_Fig2.png}}}$
\caption{Left: The 95\% CL upper limits on the mixing parameter $|V_{\mu N}|^2$ between a heavy neutrino $N$ and a muon neutrino in the mass range 5 to 50 GeV for same charge and opposite charge muon final states with and without lifetime corrections~\cite{LHCb:2020wxx}.
Right: Reconstructed mass distribution for the vector-like quark $B$ from a CMS search, comparing data, SM background estimation and several simulated signal models~\cite{CMS:2020ttz}. 
%Right: The 95\% CL upper limits on the cross section times branching ratio of the $gg \rightarrow H \rightarrow \mu\tau$ process from the LHCb search~\cite{LHCb:2018ukt}.
}
\label{fig:exo1}
\end{figure}

A different set of searches explored heavy neutral leptons.  A CMS search motivated by the left-right symmetric models studied the right-handed $W$ boson $W_R$ decaying into a heavy neutrino $N$ and a SM lepton~\cite{CMS-PAS-EXO-20-002}. The search was based on a final state with 2 same flavor leptons and 2 jets.  Cases for both Lorentz boosted and non-boosted $N$ were studied, where a lepton subjet fraction quantity was used to verify boosted $N$ decays.  The analysis looked for an excess in the invariant mass of 2 jets and 2 leptons ($m_{jj\ell\ell}$) and found no discrepancy from the SM.  For an $N$ mass of 0.2 TeV, ${W_R}$ mass is excluded at 95\% CL up to 4.7 (4.8) and 5.0 (5.4) TeV for the electron and muon channels, respectively. This analysis provided the most stringent limits on $m_{W_R}$ to date.  A different study by LHCb searched for the production of a heavy neutrino $N$ via mixing with an SM neutrino from the decay of a W boson and the subsequent semileptonic decay of $N$ into a lepton and two quarks~\cite{LHCb:2020wxx}. The search was performed in the two muons plus a jet final state, using a 3.0 fb$^{-1}$ of proton-proton collision data at center-of-mass energies of 7 and 8 TeV.  The search looked for an excess in the dimuon plus jet invariant mass distribution.  Figure~\ref{fig:exo1} (left) shows the 95\% CL observed upper limits on the mixing parameter $|V_{\mu N}|^2$ between $N$ and a muon neutrino in the $N$ mass range 5 to 50 GeV for same-sign and opposite-sign muon final states with and without lifetime corrections, for lepton number conserving ($\sim 10^{-3}$) and lepton number violating ($\sim 10^{-4}$) decays of $N$.

\begin{figure}[htp!]
\centering
$\vcenter{\hbox{\includegraphics[width=0.43\textwidth]{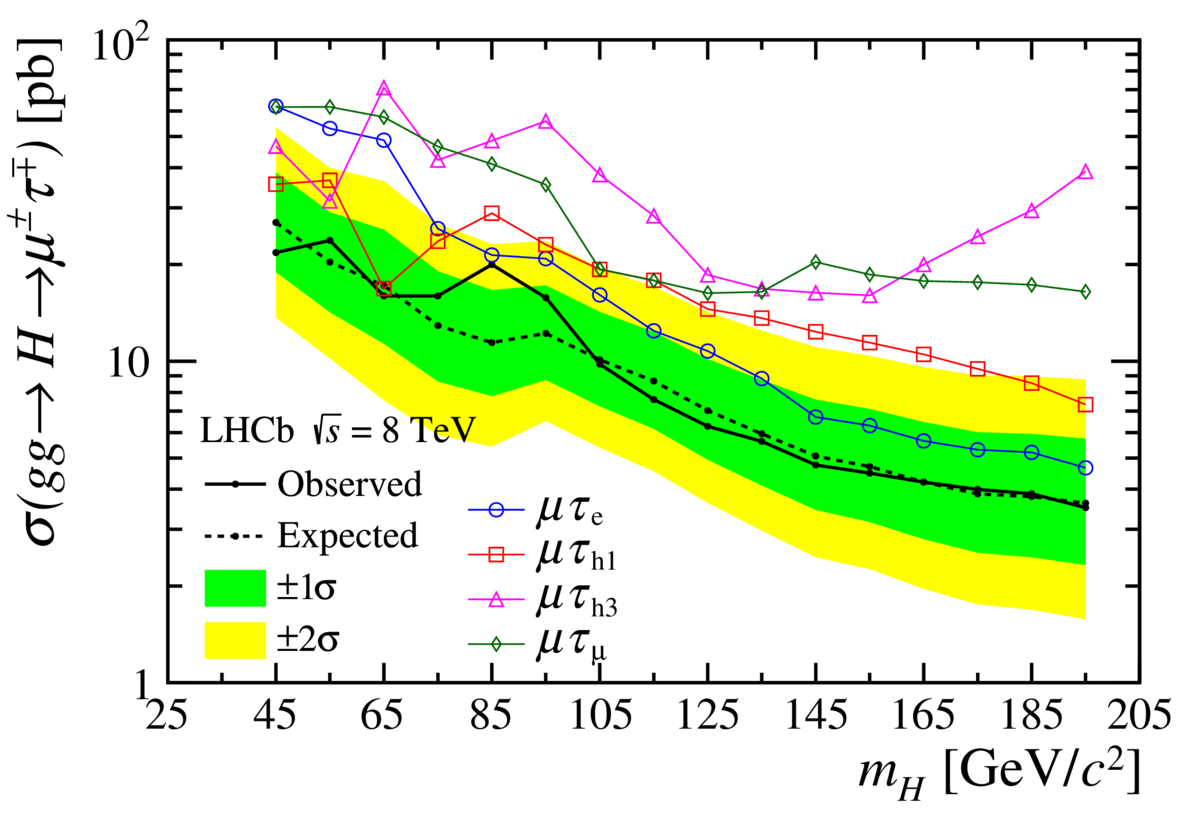}}}$
%$\vcenter{\hbox{\includegraphics[width=0.37\textwidth]{figures/CMS-B2G-19-005_Figure_009-b.png}}}$
\hspace{0.5cm}
$\vcenter{\hbox{\includegraphics[width=0.45\textwidth]{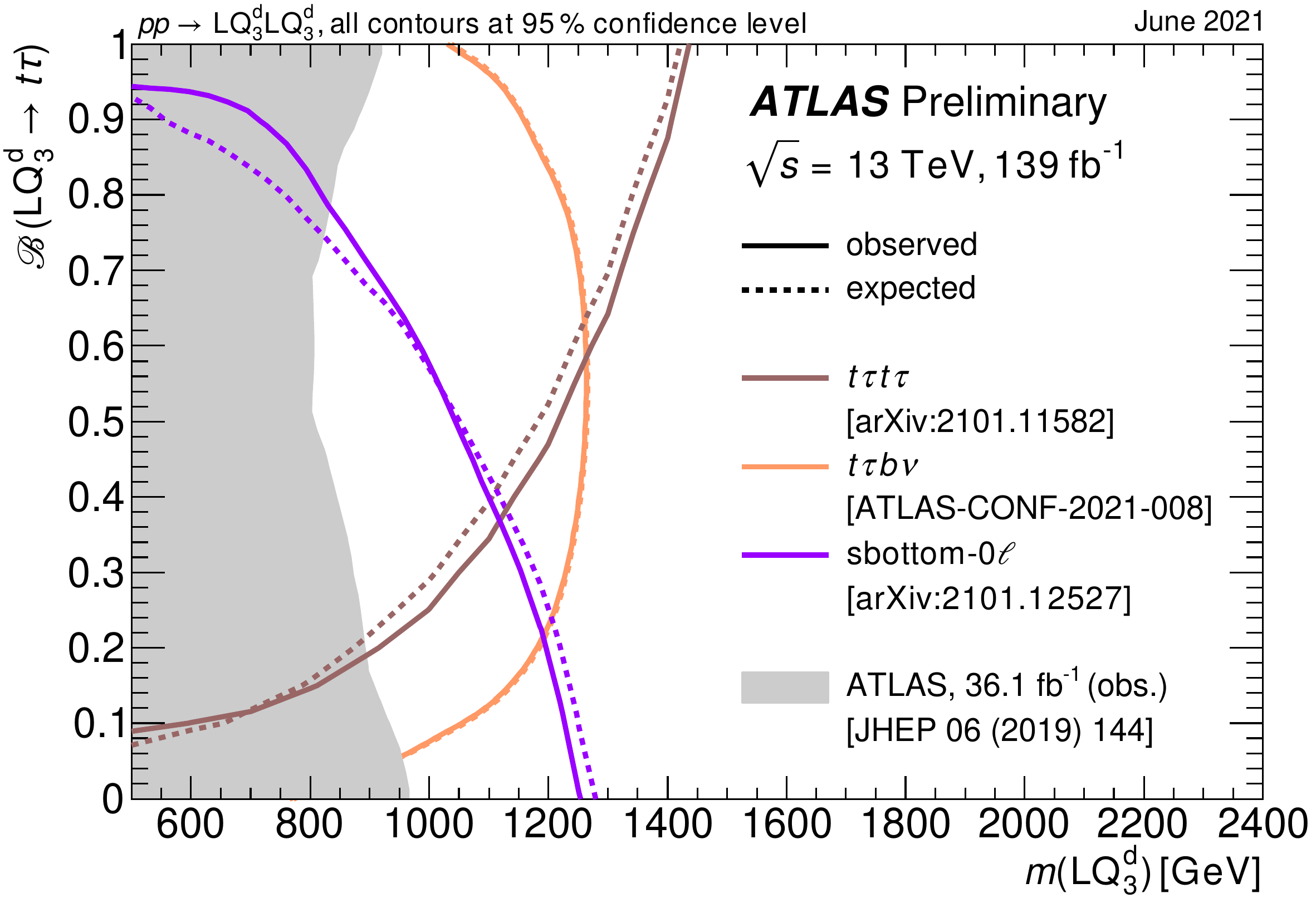}}}$
\caption{
Left: The 95\% CL upper limits on the cross section times branching ratio of the $gg \rightarrow H \rightarrow \mu\tau$ process from the LHCb search~\cite{LHCb:2018ukt}.
%Left: Reconstructed mass distribution for the vector-like quark $B$ from a CMS search, comparing data, SM background estimation and several simulated signal models~\cite{CMS:2020ttz}. 
Right: Exclusion contours at the 95\% CL from ATLAS Run 2 searches for pair-produced scalar third-generation down-type leptoquarks with decays $LQ_3^d \rightarrow b\nu / t\tau$, as a function of the leptoquark mass and the branching fraction $BR(LQ_3^d \rightarrow t\tau)$~\cite{ATL-PHYS-PUB-2021-017}.
}
\label{fig:exo2}
\end{figure}

The next class of searches look for heavy vector-like quarks.  A CMS study looked for the pair production of bottom-type vector-like quarks $B$ decaying as $B \rightarrow bZ$ and $B \rightarrow bH$ in final states with 4,5,6 jets plus b-quark jets~\cite{CMS:2020ttz}.  Different jet multiplicity categories were used to account for the fact that Higgs or Z boson decays can produce either two distinct jets or, if highly Lorentz boosted, a single merged jet. The $B$ mass was reconstructed via a $\chi^2$ metric, and its distribution was observed to be consistent with the SM as shown in Figure~\ref{fig:exo2} (left).  The analysis excluded $B$ masses up to 1.5 TeV.  Similarly an ATLAS study searched for the single production of a 3rd generation up-type vector-like quark $T$ decaying as $T\rightarrow Ht$ or $T\rightarrow Zt$  in final states containing a single lepton with multiple jets and b-jets. The presence of boosted heavy resonances in the event is exploited to discriminate the signal from the SM backgrounds~\cite{ATLAS-CONF-2021-040}.  The analysis looked for excesses in the distribution of scalar sum of object transverse momenta and found compatibility with the SM.  Upper limits are set at 95\% CL on the production cross section of T quarks in the individual decay channels. The results were interpreted in benchmark scenarios to set limits on the mass and universal coupling strength ($\kappa$) of the vector-like quark. For singlet $T$ quarks, $\kappa$ values above 0.5 were excluded for all masses below 1.8 TeV. At a mass of 1.6 TeV, $\kappa$ values as low as 0.41 were excluded.  

BSM searches are not limited to looking for new particles.  Some searches probe processes involving SM particles that are not predicted by the SM.  One such case is the searches for lepton flavor violating decays.  ATLAS performed a search for the charged lepton flavor violating process $Z \rightarrow e\mu$~\cite{ATLAS-CONF-2021-042} .  The search looked for an excess in the $e\mu$ invariant mass distribution near the $Z$ boson mass. A boosted decision tree was used to enhance signal over backgrounds.  No excess was observed, and an upper limit was placed on the branching fraction $BR(Z \rightarrow e\mu) < 3.04 \times 10^{-7}$ at 95\% CL. %calculated via a ratio to the average of observed yields of $ee$ and $\mu\mu$ events to significantly reduce systematic uncertainties.  
Similarly, an LHCb study looked for the lepton flavor violating process $H \rightarrow \mu\tau$ using 2 fb$^{-1}$ of 8 TeV proton-proton collision data~\cite{LHCb:2018ukt}. The $\tau$ leptons were reconstructed in both leptonic and hadronic decay channels. Figure~\ref{fig:exo1} (right) shows the 95\% CL upper limits on $\mathrm{\sigma \times BR}$ of the $gg \rightarrow H \rightarrow \mu\tau$ process. The exclusion ranges from 22 pb for a Higgs boson mass of 45 GeV to 4 pb for a mass of 195 GeV.

Turning to new bosons, one set of searches target leptoquarks (LQs), which are new scalar or vector bosons with simultaneous couplings to leptons and quarks.  Third generation LQs with large couplings can explain BaBar, Belle and LHCb flavor anomalies. At the LHC energies, LQs can be produced via pair production, single production or nonresonant production, but the current Run 2 searches focus on pair production.  Figure~\ref{fig:exo2} (right) shows the exclusion contours at 95\% CL from ATLAS Run 2 searches for pair-produced scalar third-generation down-type leptoquarks with decays $LQ_3^d \rightarrow b\nu / t\tau$, as a function of the leptoquark mass and the branching ratio $BR(LQ_3^d \rightarrow t\tau)$~\cite{ATLAS:2021oiz, ATLAS:2021jyv}. Moreover, an HL-LHC study in the $LQ \rightarrow \tau\mu$ and $LQ \rightarrow b\tau$ decays showed that the reach to LQ mass will improve on average by 500 GeV with respect to Run 2~\cite{CidVidal:2018eel}.  

BSM theories with enhanced symmetries predict new gauge bosons, W' and Z', which are mediators of new charged and neutral vector currents. These gauge bosons appear as resonances that can be investigated by reconstructing them from their decay products and looking for excesses in the reconstructed invariant mass or transverse mass distributions.  So far none of these searches observed a significant deviation from the SM expectation.  A recent ATLAS search looked for heavy W' in decays to $tb$ in fully hadronic final states~\cite{ATLAS-CONF-2021-043}.  It employed deep neural network-based top jet tagging and b-jet tagging to identify and discriminate signal events.  The results were interpreted in terms of a leptophobic W' with right-handed chirality, where the W' masses were excluded up to 4.4 TeV.  On the other hand, an HL-LHC search for $W' \rightarrow e\nu/\mu\nu$ predicted in the sequential standard model performed in the lepton plus $E_T^{miss}$ final state found that the HL-LHC conditions would exclude W' masses up to $\sim 8$~TeV, improving sensitivity form the corresponding Run 2 analysis by $\sim 2$ TeV~\cite{CidVidal:2018eel}.

\begin{figure}[htp!]
\centering
$\vcenter{\hbox{\includegraphics[width=0.45\textwidth]{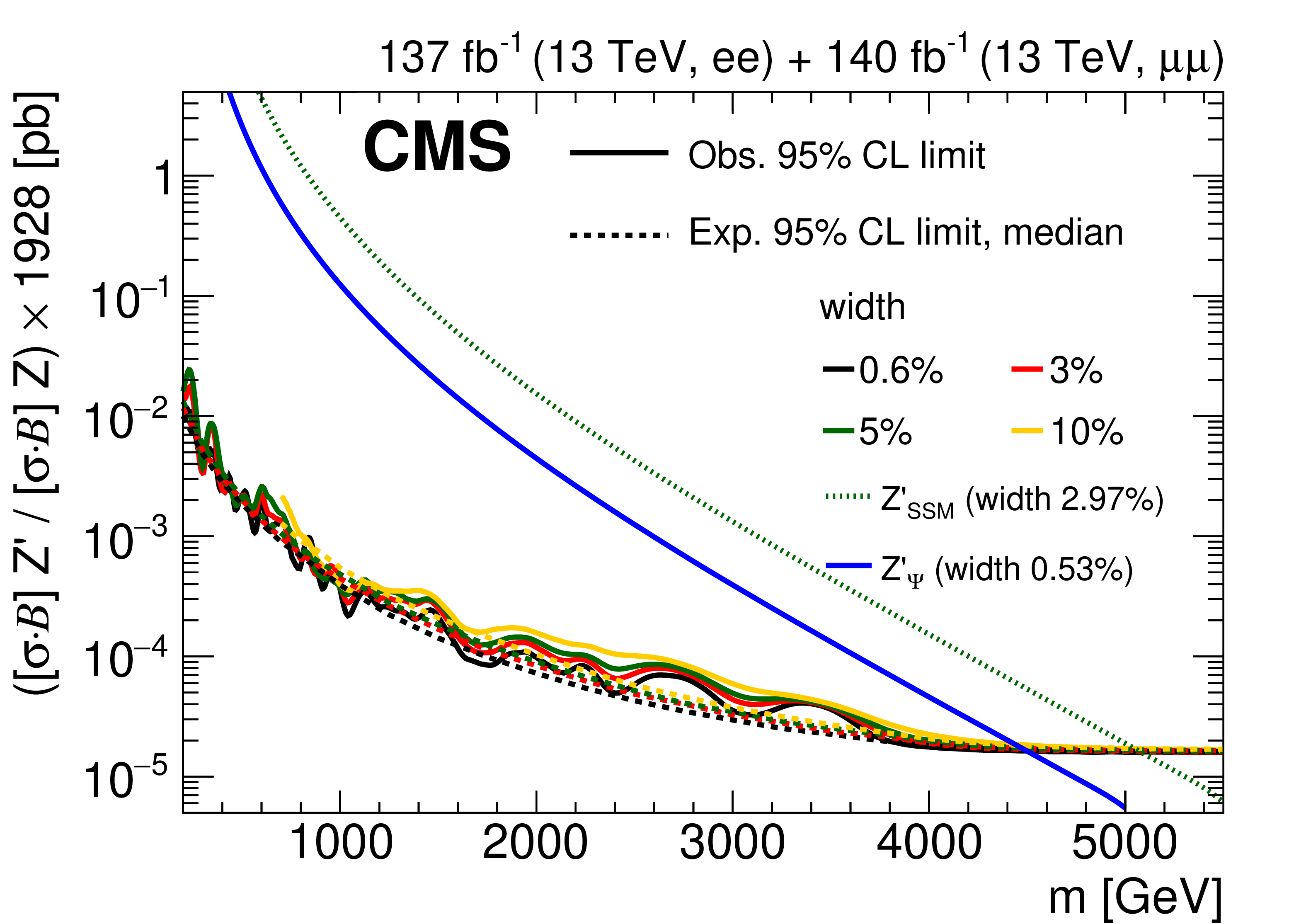}}}$
\hspace{0.5cm}
$\vcenter{\hbox{\includegraphics[width=0.37\textwidth]{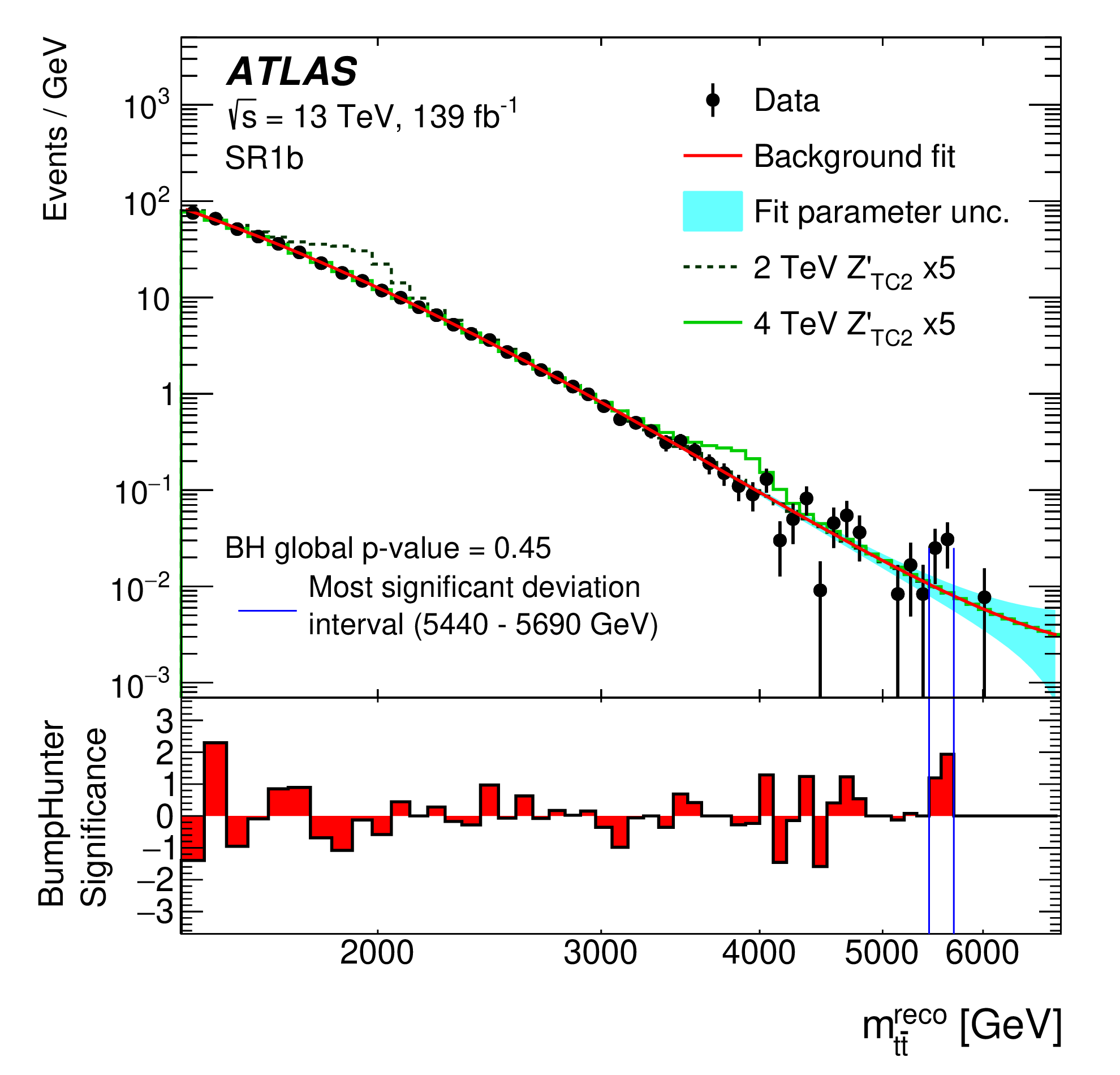}}}$
\caption{
Left: CMS upper limits at 95\% CL on the ratio of $\mathrm{\sigma \times BR}$ in a dilepton channel of a new resonance with intrinsic width up to 10\% of the SM Z boson width~\cite{CMS:2021ctt}. 
Right: Reconstructed $t\bar{t}$ resonance mass along with a background fit and distributions for several signal model points from the ATLAS $t\bar{t}$ resonance search~\cite{ATLAS:2020lks}.
}
\label{fig:exo3}
\end{figure}

Another resonance search from CMS investigated high mass spin-1 resonances such as Z', and spin-2 BSM resonances such as Randall-Sundrum gravitons, decaying to dilepton final states $ee$ or $\mu\mu$~\cite{CMS:2021ctt}.  These models are motivated by grand unified theories such as the left-right symmetric models.    
Upper limits at 95\% CL were set on the ratio of the product of $\sigma \times BR$ of a new resonance with an intrinsic width of up to 10\% to that of the SM Z boson, as shown in Figure~\ref{fig:exo3} (left).  The limits are interpreted in the context of a sequential SM (SSM) and a superstring-inspired model that predict spin-1 resonances. Lower mass limits of 5.15 (4.56) TeV are set in the $Z'_{SSM}$ and $Z'_\phi$ resonances.  Other model-dependent limits and limits on spin-2 resonances were also reported.  Additionally, an HL-LHC search for $Z' \rightarrow ee/\mu\mu$ predicted in the sequential standard model, performed in the dilepton final state found that the HL-LHC conditions would exclude Z' masses up to $\sim 6$~TeV, improving sensitivity form the corresponding Run 2 analysis by $\sim 2$ TeV~\cite{CidVidal:2018eel}.

Other studies investigated heavy BSM resonances decaying to vector bosons or the Higgs boson.  One CMS study searched for heavy resonances $X$ decaying to $Z(\nu\nu)V(qq)$, where $V$ can be a W or a Z boson~\cite{CMS:2021itu}.  Both Z and V are considered to have a high Lorentz boost, and result in a high $E_T^{miss}$ and a merged dijet through their decays, respectively.  Some signal categories also require the presence of high momentum jets in the forward region to identify production through weak vector boson fusion. Signal is extracted by a fit to the transverse mass calculated from $E_T^{miss}$ and the merged dijet system, where the latter is identified as a large radius jet with substructure.  The analysis reported lower limits on various resonance masses, excluding ggF produced radions up to 3.0 TeV, Drell-Yan produced W' bosons up to 4.0 TeV, and ggF produced gravitons up to 1.2 TeV.  In addition, upper observed limits on the product of the VBF production cross section and $X\rightarrow Z+W/Z$ branching fraction range between 0.2 and 20 (0.3 and 30) fb.  A second search performed by CMS looked for heavy resonances $X$ decaying to $W(\ell \nu)V(qq)$ or $W(\ell \nu)H(qq)$ pairs, where $V$ can be a W or a Z boson~\cite{CMS:2021klu}.  Once again, the V or H decaying to a dijet pair leads to a merged, large radius due to high Lorentz boost.  Analysis is performed in final states with a lepton, $E_T^{miss}$, a merged, large radius jet, and for some search regions, forward jets to address VBF signatures.  Signal is extracted in a 2-dimensional fit to the merged jet invariant mass versus reconstructed diboson transverse mass.  The analysis excluded ggF produced spin-2 bulk gravitons decaying to WW up to 1.8 TeV, spin-1 Drell-Yan-produced Z' decaying to WW up to 3.9 TeV, W' bosons decaying to WZ and WH up to 3.9 and 4.0 TeV, respectively, and ggF produced spin-0 bulk radions up to 3.1 TeV. 

Finally, heavy resonances can be explored in their decays to top pairs.  An ATLAS search was performed using events consistent with pair production of high-transverse-momentum top quarks and their subsequent decays into fully hadronic final states~\cite{ATLAS:2020lks}.  The analysis was optimized for resonances decaying into a $t\bar{t}$ pair with mass above 1.4 TeV, exploiting a dedicated multivariate technique with jet substructure to identify hadronically decaying top quarks using large-radius jets. Signal was extracted in a fit to reconstructed $t\bar{t}$ invariant mass, as shown in Figure~\ref{fig:exo3} (right).  Limits were set on the production cross section times branching fraction for a Z' boson within the framework of a topcolor-assisted-technicolor model. The Z' boson masses below 3.9 and 4.7 TeV were excluded at 95\% CL for the decay widths of 1\% and 3\%, respectively.  Additionally, CMS performed an HL-LHC sensitivity study for $t\bar{t}$ resonances in the fully hadronic and single lepton final states, which predict the expected exclusion on the mass of a Randall-Sundrum gluon to be up to 6.6 TeV~\cite{CidVidal:2018eel}.

\section{Summary}

The ATLAS, CMS and LHCb experiments performed hundreds of searches for beyond the standard model physics with the LHC Run 2 data, covering a large variety of models and signatures.  This report presented recent examples from generic searches, searches for supersymmetry, searches for extended Higgs sectors and searches for new fermions and bosons.  None of those searches so far have discovered a significant deviation from the standard model expectations.  The analysis results have been interpreted in terms of various models, and lower limits on new particle masses or upper limits on cross sections have been reported.  The quest is ongoing with continuous design of more searches, which use ever more refined analysis techniques.  In the meanwhile, preparations are underway for LHC Run 3, and finally for the High-Luminosity LHC, which will offer unprecedented prospects for new physics searches.  Highlights from these prospects were also presented. More future studies are in progress, exploring innovative search methods that would allow us to make the most of this exciting new era.  

\section*{Acknowledgments}

I thank all colleagues in the ATLAS, CMS and LHCb experiments who have contributed to the studies described in the report.  SS is supported by the Basic Science Research Program through the National Research Foundation of Korea (NRF) funded by the Ministry of Education under the contracts  NRF-2008-00460 and NRF-2018R1A6A1A06024970.

\section*{References}

\bibliographystyle{unsrt}
\bibliography{SekmenBlois2021}

\end{document}